\documentclass[%
 reprint,
 superscriptaddress,
 amsmath,amssymb,
 aps,
]{revtex4-2}
\usepackage{xcolor}
\usepackage{graphicx}
\usepackage{dcolumn}
\usepackage{bm}

\begin{document}
\preprint{APS/123-QED}

\title{Direct probing of strong magnon-photon coupling in a planar geometry}

\author{Mojtaba Taghipour Kaffash}%
\author{Dinesh Wagle}
\author{Anish Rai}
\affiliation{Department of Physics and Astronomy, University of Delaware, Newark, Delaware 19716, United States}
\author{Thomas Meyer}
\affiliation{THATec Innovation GmbH, D-67059 Ludwigshafen,
Germany}
\author{John Q. Xiao}
\author{M. Benjamin Jungfleisch}\email{mbj@udel.edu}
\affiliation{Department of Physics and Astronomy, University of Delaware, Newark, Delaware 19716, United States}

\date{\today}

\begin{abstract}
We demonstrate direct probing of strong magnon-photon coupling using Brillouin light scattering spectroscopy in a planar geometry. The magnonic hybrid system comprises a split-ring resonator loaded with epitaxial yttrium iron garnet thin films of 200 nm and 2.46 $\mu$m thickness. The Brillouin light scattering measurements are combined with 
microwave spectroscopy measurements where both biasing magnetic field and microwave excitation frequency are varied. The cooperativity for the 200 nm-thick YIG films is 4.5, and larger cooperativity of 137.4 is found for the 2.46 $\mu$m-thick YIG film. We show that Brillouin light scattering is advantageous for probing the magnonic character of magnon-photon polaritons, while microwave absorption is more sensitive to the photonic character of the hybrid excitation. A miniaturized, planar device design is imperative for the potential integration of magnonic hybrid systems in future coherent information technologies, and our results are a first stepping stone in this regard. Furthermore, successfully detecting the magnonic hybrid excitation by Brillouin light scattering is an essential step for the up-conversion of quantum signals from the optical to the microwave regime in hybrid quantum systems.

\end{abstract}

\maketitle

The emergent properties of hybrid systems are promising for a wide range of quantum information applications. In particular, light-matter interaction has been at the forefront of contemporary studies on hybrid quantum systems. To this end, hybrid magnonic systems based on the coupling of magnons, the elementary excitations of magnetic media, and photons have gained increased attention \cite{Hu_2015, Harder_2018, Li_Review_2021,Bhoi_2019}. Magnons display a highly tunable dispersion, while they can be used for coherent up- and down conversion between microwave and optical photons \cite{Hisatomi_2016,XZhang_PRL_2016,Haigh_2016,Osada_2018,Klingler_2016}. In addition, magnons can serve in quantum memory applications owing to their collective behavior and robustness \cite{Schultheiss_2019}.

A critical requirement for coherent information transfer based on magnons is a high cooperativity, which means that the coupling between the two disparate types of excitations, i.e., the photonic and the magnonic subsystems, exceeds the loss rates of either subsystem. This is known as the strong coupling regime in the language of quantum information. In this strong coupling regime, information 
can be efficiently exchanged, potentially enabling efficient transduction applications. Another prerequisite for large scale quantum information processing and transfer applications is the conversion between optical and microwave frequencies. Previous microwave-to-optical transduction studies based on ferromagnets either employed Brillouin scattering of optical whispering gallery modes by magnetostatic modes \cite{XZhang_PRL_2016,Haigh_2016,Osada_2018} or coupling of the microwave field through a cavity mode concomitant with the coupling of the optical field through the Kittel mode via Faraday and inverse Faraday effects \cite{Hisatomi_2016}. Most of these prior works relied on macroscopic samples made of bulk yttrium iron garnet (YIG) crystals. Utilizing YIG is advantageous as it has a large spin density and narrow linewidth \cite{Tabuchi_2014,Bhoi_2014,Harder_PRL2018,Bai_2015}. However, scalable on-chip solutions require device miniaturization. Therefore, planar microwave resonators are advantageous for building hybrid magnonic networks and circuits \cite{Castel2017}. They offer great flexibility in terms of circuit design; they are compatible with lithographic fabrication processes and the prevalent complementary metal-oxide-semiconductor (CMOS) platform \cite{chumak2021roadmap}. Furthermore, planar microwave resonators typically have a smaller effective volume than their three-dimensional counterparts and can provide an enhanced coupling with magnetic dipoles \cite{Li_PRL2019,Hou_2019}. In addition, they potentially simplify the integration of optical components \cite{Klingler_2016} enabling simplified optomagnonic device concepts.

\begin{figure}[ht]
\centering
\includegraphics[width=0.9\columnwidth]{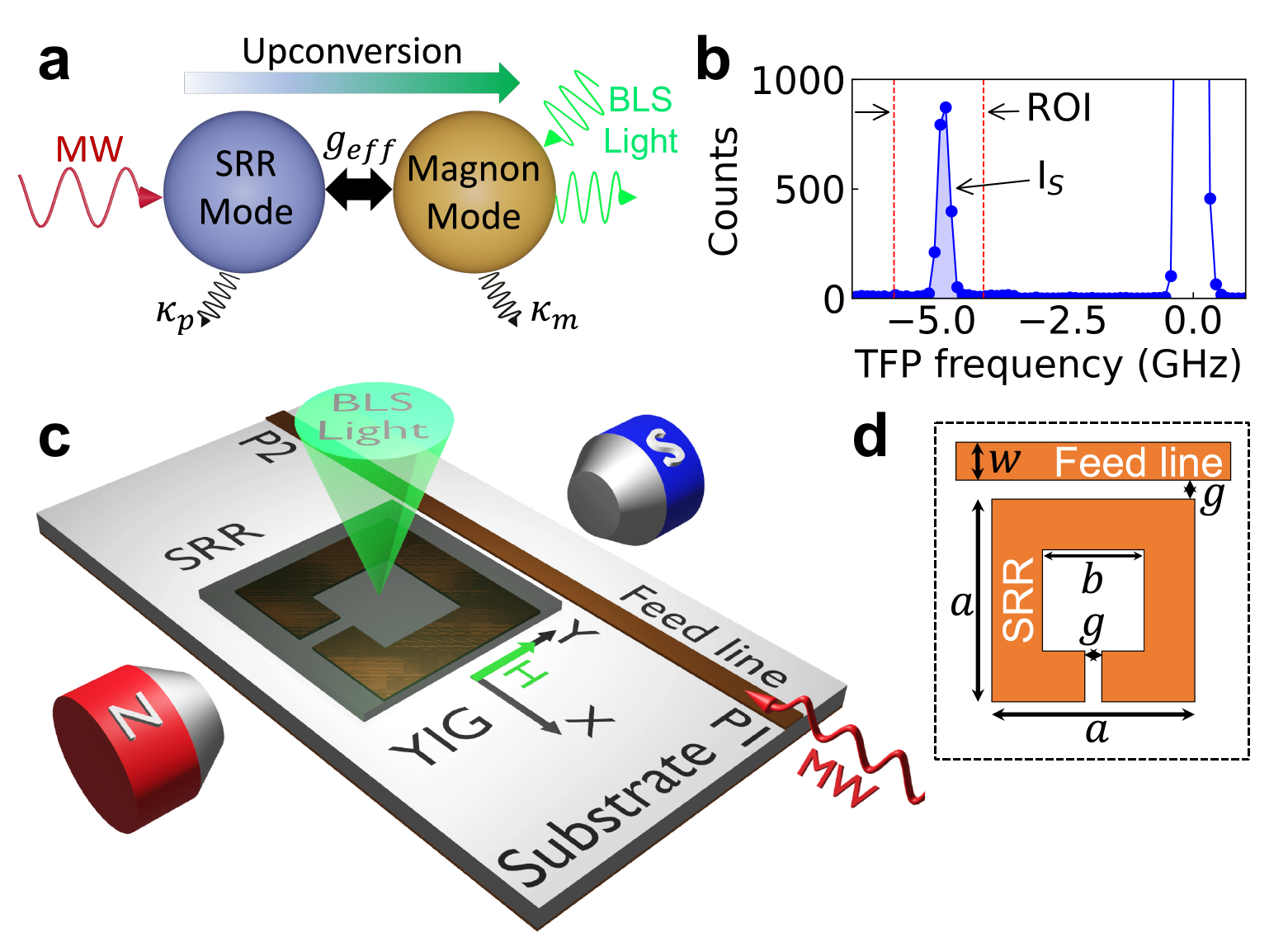}
\caption{(a) Schematic illustration of the coupling process between the microwave photon (MW) mode of the split-ring resonator (SRR) with the magnon mode of the YIG film, where $\kappa_\mathrm{p}$ and $\kappa_\mathrm{m}$ are the dissipation rates of microwave photon and magnon, respectively, and $g_\mathrm{eff}$ is their mutual coupling strength. Microwave-to-optical up-conversion is achieved by coupling the incident microwave photons via the the SRR to the magnon mode that interacts with the BLS laser photons. (b) A typical BLS spectrum with the Rayleigh peak at 0~GHz and the Stokes signal at around -5~GHz. The vertical red dashed lines show the region of interest (ROI). (c) Experimental setup: The resonator consists of a square SRR patterned next to the microwave feed line. The YIG film is placed on the top of the SRR. An external biasing magnetic field (in $y$-direction) magnetizes the sample during the 
BLS and MW measurements. The probing BLS beam is focused onto the surface of the YIG film. (d) Top view of the SRR with the dimensions as defined in the text.
} 
\label{fig:schem}
\end{figure}

Here, we demonstrate coherent microwave-to-optical up-conversion using strong magnon-photon coupling in a split-ring resonator/YIG thin film hybrid circuit. We directly probe the coupling in YIG films of 200 nm and 2.46 $\mu$m thickness by conventional and microfocused Brillouin light scattering (BLS) spectroscopy and compare these optical results to microwave absorption measurements. Clear avoided level crossings are observed evidencing the hybridization of the magnon and microwave photon modes in the strong coupling regime. In addition, we identify contributions of higher order magneto-static surface spin waves. The cooperativity for the 200 nm-thick YIG films is 4.5 and 137.4 for the 2.46~$\mu$m-thick YIG film. On the one hand, we find that BLS is advantageous for probing the magnonic character of magnon-photon polaritons, while microwave absorption is found to be more sensitive to the photonic character. On the other hand, detecting the magnonic hybrid excitation by Brillouin light scattering demonstrates a coherent conversion of microwave to optical photons. 


The coherent microwave-to-optical up-conversion process based on the strong magnon-photon coupling is illustrated in Fig.~\ref{fig:schem}(a). The magnonic hybrid system comprises a split-ring resonator (SRR) loaded with epitaxial YIG thin films. The microwave photons interact with the SRR mode that exhibits a dissipation rate of $\kappa_\mathrm{p}$ at its resonance frequency. The SRR mode couples with the magnon mode of the YIG sample with a coupling constant of $g_\mathrm{eff}$, while the YIG sample dissipates its energy at the rate $\kappa_\mathrm{m}$. Finally, the excited magnons interact and couple with the incident BLS probe beam.

The up-conversion process is realized by two separate sets of measurements: in-plane magnetic field dependent microwave (MW) absorption measurements and  BLS (both microfocused and conventional) of RF driven magnetization dynamics. A typical BLS spectrum is shown in Fig.~\ref{fig:schem}(b), where the region of interest (ROI) is limited to the frequency range of hybrid excitation (here: Stokes peak, I$_S$). The elastically scattered light is centered at 0~GHz. {The probing BLS laser beam is focused on the sample surface; therefore, we detect magnons modes only in the top layer \cite{Kargar_2021}, while both the top and bottom layers contribute in the MW absorption measurements. However, since each sample is grown under the same fabrication procedure, similar properties are expected from each YIG-film layer in each sample}. Figure~\ref{fig:schem}(c) depicts the experimental configuration consisting of the square SRR in the vicinity of an MW feed line loaded with a YIG sample placed on the top and in the presence of a biasing in-plane magnetic field applied along the $y$-axis. For the MW absorption measurement, a vector network analyzer (VNA) is used to record the field-dependent transmission parameter S$_{12}$ with 
 {an output} power of +13~dBm connected to P1 and P2. We use a continuous single-mode 532-nm wavelength laser for the BLS measurements that is focused on the YIG film's surface [see Fig.~\ref{fig:schem}(c)]. A MW generator provides the a RF signal to the feed line (P1) with 
 {output} powers of +20~dBm for microfocused BLS and +27~dBm for conventional BLS measurements. The BLS process can be described by the inelastic scattering of laser photons with magnons \cite{Jungfleisch_2020}. Since this process is energy and momentum conserving, inelastically scattered photons carry information about the probed magnons \cite{madami_2012_book}, which we analyze using a high-contrast tandem Fabry-P\'erot interferometer. Two different objective lenses are used for the BLS measurements: for the microfocused measurements, a high-numerical-aperture (NA = 0.75) objective lens with a working distance of 4~mm is used, while a lens with a focal lens of 40~mm and a diameter of 1~inch is used for the conventional measurement setup.

\begin{figure}[t]
\centering
\includegraphics[width=0.99\columnwidth]{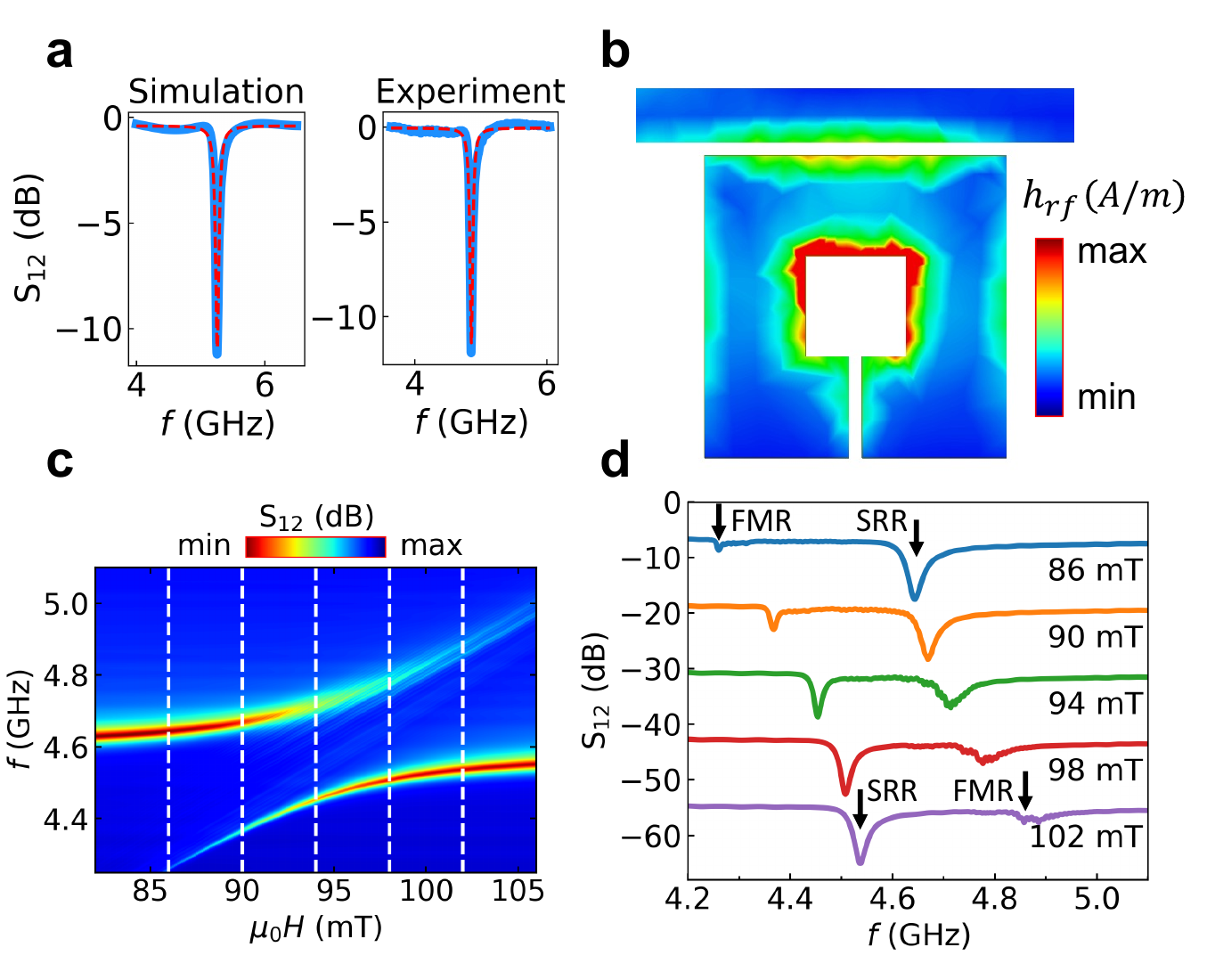}
\caption{(a) SRR resonance obtained by HFSS simulations ($Q_\mathrm{sim}=83.1$) and corresponding experimentally realized resonance ($Q_\mathrm{exp}=94.0$). Data shown in blue, corresponding fits are shown by in red dashed lines. (b) $RF$ magnetic field ($h_\mathrm{rf}$) distribution obtained by HFSS simulations. (c) MW absorption measurements of the magnon-photon hybridization (here, YIG film thickness: $2.46~\mu$m), where the false color represents the S$_{12}$ transmission parameter. (d) S$_{12}$ transmission parameter versus frequency $f$ at selected biasing magnetic fields as shown by white dashed lines in (c). 
} 
\label{fig:hfss}
\end{figure}

\begin{figure*}[t]
\centering
\includegraphics[width=1.0\textwidth]{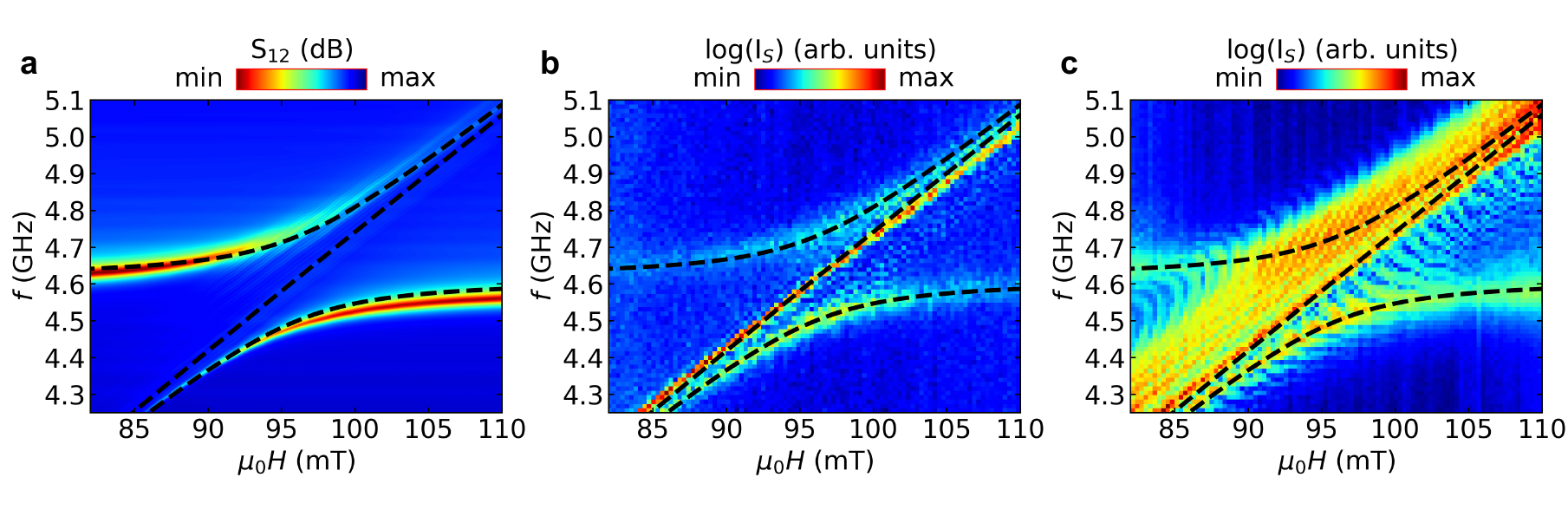}
\caption{False color-coded spectra of the magnon-photon hybridization of the 2.46~$\mu$m-thick YIG film. Results obtained by (a) microwave absorption measurements, where $S_{12}$ is plotted versus $f$ and $\mu_0H$, (b) microfocused BLS spectroscopy, and (c) conventional BLS spectroscopy. In the BLS measurements, the Stokes peak [compare to Fig.~\ref{fig:schem}(b)] is plotted in logarithmic scale versus $f$ and $\mu_0H$. The black dashed lines are the fits to Eqs.~(\ref{eq:hybrid}) and (\ref{eq:kittel}).
}
\label{fig:2um}
\end{figure*}


We designed and optimized the SRR via ANSYS HFSS to exhibit a resonance 
($f_0$) at 5.1~GHz, which agrees with the experimentally observed result (4.9~GHz) as shown in Fig.~\ref{fig:hfss}(a). Figure~\ref{fig:schem}(d) illustrates the top view of the SRR with the following dimensions: the SRR's outer and inner widths of $a=4.5~$mm and $b=1.5~$mm, the gap between the SRR and the feed line $g=0.2~$mm, and the feed line's width of $w=0.4~$mm. The SRR is fabricated by etching one side of Rogers RO3010 laminate with a dielectric constant of 10.20${\pm}$0.30 and copper thickness of 35~${\mu}$m that is coated on both sides of the substrate. By fitting the resonance data to a Lorentzian function with full-width at half maximum (FWHM), we determine the quality factor ($Q=f_0/{\Delta}f_\mathrm{FWHM}$) of the resonator to be $Q_\mathrm{sim}$=83.1 for the simulation and $Q_\mathrm{exp}$=94.0 for the experiment [shown with the red dashed lines in Fig.~\ref{fig:hfss}(a)]. The 2D profile of the modeled RF-magnetic field $h_\mathrm{rf}$ on resonance is shown in Fig.~\ref{fig:hfss}(b). $h_\mathrm{rf}$ is the most intense and uniform at the center of the SRR. The SRR is loaded with low-loss YIG films placed on the top of the center of the SRR [for details on broadband ferromagnetic resonance measurements we refer to the supplemental material (SM)]. We compare the results of two YIG-film thicknesses: the lateral dimensions of the 2.46~$\mu$m thick square-shaped sample is 5.3~mm~$\times$~5.3~mm, while the 200~nm thick-sample is parallelogram-shaped with a base and height of 10~mm and 7.5~mm, respectively. Both samples are grown on 500~$\mu$m-thick gadolinium gallium garnet substrates by liquid phase epitaxy on both sides of the substrates.

Fig.~\ref{fig:hfss}(c) shows a typical false color-coded microwave absorption spectrum of the magnon-photon hybridization (here, YIG film thickness: $2.46~\mu$m), where the color represents the transmission parameter. In the field/frequency region where the uncoupled photon and the magnon modes would cross, we observe the behavior of an effective two-level system, where the two disparate subsystems couple electromagnetically with the coupling strength $g_\mathrm{eff}$. The coupling is quantified by the cooperativity $C=g_\mathrm{eff}^2/\kappa_\mathrm{m}\kappa_\mathrm{p}$. The mode coupling lies in the strong regime if $g_\mathrm{eff}$ is larger than the loss rate of YIG, $\kappa_\mathrm{m}$, and the SRR, $\kappa_\mathrm{p}$, respectively \cite{zhang_2014_PRL}; thus, $C>1$.
This is shown more in detail in Fig.~\ref{fig:hfss}(d), where S$_{12}$ is plotted versus $f$ for different fields from 86 to 102~mT close to the avoided crossing as indicated by white dashed lines in Fig.~\ref{fig:hfss}(c). {At high fields (e.g., at 102~mT), the higher frequency mode (FMR mode) has a lower intensity than the lower frequency mode (SRR mode). By sweeping the field from higher to lower values, the FMR mode approaches the SRR mode. In this transition regime, the modes switch the magnitude of their intensities: at 94~mT, both modes have the same intensity, and the frequency gap between them is almost minimum. Further decreasing the field magnitude to 86~mT results in the modes switching their intensities and moving apart. This behavior describes an avoided level crossing indicative of the formation of magnon-photon polaritons \cite{rameshti_2021_arXiv, Bhoi_2020_book}.}

We model the photon-magnon hybridization using a coupled two harmonic oscillator model with $f_{\pm}$ representing the hybridized mode frequencies: 
\begin{equation}\label{eq:hybrid}
    f_{\pm}=\frac{f_\mathrm{SRR}+f_\mathrm{FMR}}{2}\pm\sqrt{\left( \frac{f_\mathrm{SRR}-f_\mathrm{FMR}}{2}\right)  ^2+\left( \frac{g_\mathrm{eff}}{2}\right)^2},
\end{equation}

where $g_\mathrm{eff}$ is the coupling strength, $f_\mathrm{SRR}$ is the uncoupled SRR resonance, $f_\mathrm{FMR}$ is the ferromagnetic resonance of YIG that increases as the field is increased and is given by the Kittel formula
\begin{equation}\label{eq:kittel}
    f_\mathrm{FMR}=\frac{\gamma}{2\pi}\mu_0\sqrt{H(H+M_\mathrm{eff})},
\end{equation}

where $\gamma$ is the gyromagnetic ratio and $M_\mathrm{eff}$ is the effective magnetization (see SM). The systematic deviations of the BLS data from the FMR fitting are discussed in the SM.

The microwave absorption measurement result of the $2.46~\mu$m-thick YIG film shows a clear avoided crossing which centers at {96~mT},  Fig.~\ref{fig:2um}(a). 
As is visible from the figure, the signal is particularly strong before and after the avoided crossing ($<82$~mT and $>110$~mT). This field-independent signal is the SRR resonance mode. However, a pronounced avoided crossing is observed when the field-dependent FMR mode of YIG approaches the SRR resonance at {96 mT} leading to the formation of a hybridization. The upper and lower frequency modes and the uncoupled FMR mode are fitted to the experimental results according to Eq.~(\ref{eq:hybrid}) and Eq.~(\ref{eq:kittel}), respectively.


Using a similar field/frequency sweep range as in the microwave absorption measurements, we probe the magnon-photon hybridized state by microfocused BLS [Fig.~\ref{fig:2um}(b)]. Here, the Stokes BLS intensity in logarithmic scale is plotted. 
The field is swept from 110 to 82~mT in 0.3~mT field steps after saturating at 200~mT. The MW frequency excites the sample from 4.25 to 5.10~GHz in 12~MHz steps. The two hybridized modes are detectable, similar to the MW absorption measurements. Note that, in addition to the coupled resonances, we detect a contribution of modes directly excited by the feed line \cite{Bhoi_2014} in the BLS experiments. We will discuss these modes below.

BLS's successful detection of the strongly coupled magnon-photon state demonstrates a coherent microwave-to-optical up-conversion based on the scheme shown in  Fig.~\ref{fig:schem}(a). Interestingly, the intensity distribution detected in BLS is reverse to the MW absorption technique: BLS is more sensitive in probing the magnonic character of the magnon-photon polariton compared to MW absorption measurements shown in Fig.~\ref{fig:2um}(a), which is more sensitive to the photonic character of the hybrid excitation confirming previous reports~\cite{Klingler_2016}. 

\begin{figure}[t]
\centering
\includegraphics[width=0.8\columnwidth]{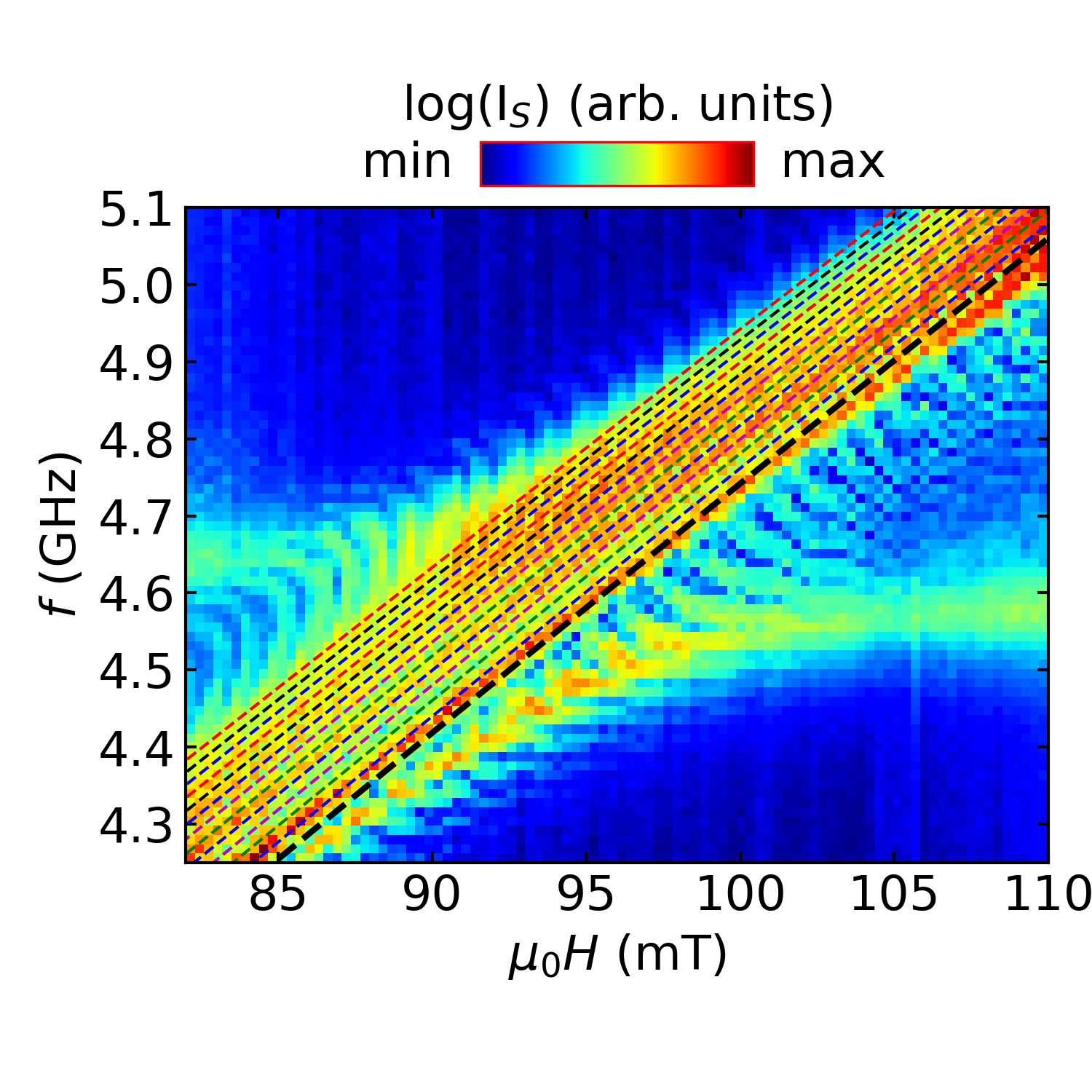}
\caption{Conventional BLS spectra of the 2.46~$\mu$m-thick YIG film. Dashed lines represent fits to Eq.~(\ref{eq:MSSW}). The black bold dashed line represents the $k=0$ ($n=0$) mode, which is the FMR mode, while the higher-lying dashed lines are MSSW modes ($n$ = 1,..., 12) with $k=n\pi/2l$.
} 
\label{fig:higher_order}
\end{figure}

\begin{figure*}[t]
\centering
\includegraphics[width=1\textwidth]{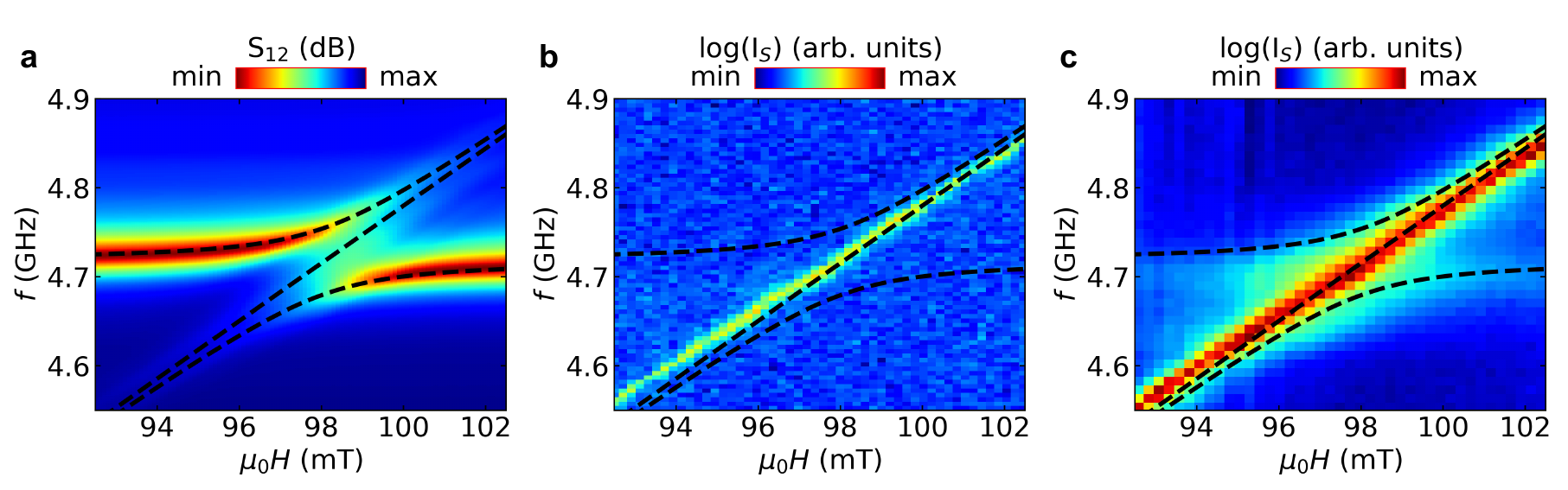}
\caption{A typical false color-coded spectrum of the magnon-photon hybridization for the 200~nm-thick YIG film using the (a) microwave absorption technique, where $S_{12}$ is plotted versus $f$ and $\mu_0H$, (b) microfocused BLS technique, and (c) BLS technique with a conventional objective lens. The dashed lines are the fitted plots to Eqs. (\ref{eq:hybrid}) and (\ref{eq:kittel}).
} 
\label{fig:200nm}
\end{figure*}
By fitting the experimental data to Eq.~(\ref{eq:hybrid}), we extract the magnon-photon coupling strength $g_\mathrm{eff}$. 
Ferromagnetic resonance measurements (SM) yield the following parameters: $\mu_0M_\mathrm{eff}=183.5$~mT and $\gamma/2\pi=28.2$~GHzT$^{-1}$, which
we use to fit the microfocused BLS results [Figs.~\ref{fig:2um}(b,c)] to Eq.~(\ref{eq:hybrid}), we obtain $g_\mathrm{eff}/2\pi=114.6$~MHz. 
By calculating the dissipation rates 
of the microwave photon ($\kappa_\mathrm{p}/2\pi=25.8$~MHz) and the magnon ($\kappa_\mathrm{m}/2\pi=3.9$~MHz), 
we can obtain a cooperativity of $C=137.4$, which fulfils the conditions $C>1$ and $g_\mathrm{eff}>\kappa_\mathrm{p},\kappa_\mathrm{m}$. 


We compare the microfocused BLS experiments to  conventional BLS measurements as is shown in Fig.~\ref{fig:2um}(c). We use a lens with a smaller numerical aperture than the objective lens used in the microfocused setup. However, the laser beam spot size is significantly larger, and hence, it covers a larger area of the YIG film leading to a stronger signal intensity. 
Due to the stronger signal strength, we are able to detect modes inaccessible by the microfocused system as further evidenced in Fig.~\ref{fig:higher_order}. 
The additional fine features revealed by the conventional BLS measurements lie in the anticrossing region parallel to the Kittel mode. These modes are due to the excitations of higher-order wavenumber spin-wave modes directly excited by the MW feed line. These modes occur at  frequencies higher than the Kittel mode for a given magnetic field and are identified as magnetostatic surface spin waves (MSSWs) that propagate in the film plane in a direction perpendicular to the applied field \cite{stenning_2013_Optica, DZhang_2017_JPD, Bhoi_2014, lambert_2015_JAP}. We model them by:
\begin{equation}\label{eq:MSSW}
f_\mathrm{MSSW} = \frac{\gamma}{2\pi}\mu_0\sqrt{H(H+M_\mathrm{eff})+M_\mathrm{eff}^2(1-e^{-2kd})/4},
\end{equation}
where $d$ is the thickness of the sample, $k = n\pi/2l$ is the spin-wave wavevector, $l$ is the length of the square-shaped sample and $n$ is the mode number with $n=0$ being the uniform Kittel mode. $\gamma/2\pi=28.2$~GHzT$^{-1}$ and $\mu_0M_\mathrm{eff}$ = 183.5~mT both of which are obtained from the fitting of the lowest lying mode, Eq.~(\ref{eq:kittel}). The dashed lines above the main modes in Fig.~\ref{fig:higher_order} shows fits of the experimental data to Eq.~(\ref{eq:MSSW}) for $n = 0, 1,..., 12$. Here, the bold dashed line represents the $k=0$ ($n=0$) mode, which is the FMR mode, while the other dashed lines are the higher-order MSSW modes ($n$ = 1,..., 12). These higher-order MSSW modes have wavevectors of $k=3.5\times10^{-3}$~rad/$\mu$m for $n=12$, which is within the detectable wavevector range of our conventional system ($k_\mathrm{max}=6.9$~rad/$\mu$m).

While most recent works on strong-magnon photon coupling utilized micrometer-thick-YIG or YIG spheres \cite{bhoi_2021_JAP, XZhang_2019_PRL, Ihn_2020_PRB, Boventer_2020_PRR}, sample miniaturization is imperative for a scalable on-chip solutions. 
In the following, we demonstrate magnon-photon coupling in a miniaturized 200~nm-thick YIG film. Using the identical 2D planar resonator as used for studying the 2.46~$\mu$m film, we observe mode anti-crossing by the microwave absorption technique as shown in Fig.~\ref{fig:200nm}(a). 
As is shown Fig.~\ref{fig:200nm}(b), we are unable to detect a sufficiently strong signal of the hybridized excitation in the microfocused measurements. Surprisingly, the Kittel mode directly excited by the feed line is significantly stronger than the magnon-photon coupled modes. However, as we switch from the microfocused to the conventional BLS setup, not only the Kittel mode becomes more intense, but also the two hybridized mode can be detected in the spectra [Fig.~\ref{fig:200nm}(c)]. 
Fits to the experimental data agree reasonably well as shown by the black dashed lines. 
From the combined optical and microwave experiments, we extract the following parameters: $\mu_0M_\mathrm{eff}=187.3$~mT, $\gamma/2\pi=28.2$~GHzT$^{-1}$, and $g_\mathrm{eff}/2\pi=37.3$~MHz. The photon and magnon dissipation rates are found to be $\kappa_\mathrm{p}/2\pi=25.8$~MHz and $\kappa_\mathrm{m}/2\pi=12.1$~MHz, respectively. Therefore, the cooperativity $C=4.5$, fulfilling both conditions $C>1$ and $g_\mathrm{eff}>\kappa_\mathrm{p},\kappa_\mathrm{m}$ and, hence, the mode hybridization is in the strong coupling regime. Higher modes similar to the ones observed in the 2.46~$\mu$m are absent in the spectra of the 200~nm film since the field/frequency separation of the higher-order modes decreases as the YIG thickness decreases \cite{kalinikos_1986_JPC}.

In summary, we showed direct probing of strong magnon-photon coupling using Brillouin light scattering spectroscopy in a planar geometry. The optical measurements are combined with 
microwave spectroscopy experiments where both biasing magnetic field and microwave excitation frequency are varied. The miniaturized YIG sample of 200 nm thickness exhibits a cooperativity of 4.5, while 2.46 $\mu$m-thick film showed a larger cooperativity of 137.4. We find that Brillouin light scattering is advantageous for probing the magnonic character of magnon-photon polaritons, while microwave absorption is more sensitive to the photonic character of the hybrid excitation. In addition, modes directly excited by the feed line significantly contribute to the optical measurements: they are detected in the gaped region between the two coupled magnon-photon modes. The detection of the magnonic hybrid excitation by Brillouin light scattering can be understood as an up-conversion mechanism of signals from the optical to the microwave regime in the magnonic hybrid systems. The planar structure presented here enables spatially-resolved imaging of magnon-photon polaritons that can serve as a platform for studying magnonics strongly coupled to microwave photons.\\


\vspace{-2ex}
\section*{Acknowledgment}
\vspace{-1ex}
We thank Prof. Matthew Doty, University of Delaware, for valuable discussions. Research supported by the U.S. Department of Energy, Office of Basic Energy Sciences, Division of Materials Sciences and Engineering under Award DE-SC0020308. The authors acknowledge the use of facilities and instrumentation supported by NSF through the University of Delaware Materials Research Science and Engineering Center, DMR-2011824.
\bibliography{apssamp}

\begin{thebibliography}{32}%
\makeatletter
\providecommand \@ifxundefined [1]{%
 \@ifx{#1\undefined}
}%
\providecommand \@ifnum [1]{%
 \ifnum #1\expandafter \@firstoftwo
 \else \expandafter \@secondoftwo
 \fi
}%
\providecommand \@ifx [1]{%
 \ifx #1\expandafter \@firstoftwo
 \else \expandafter \@secondoftwo
 \fi
}%
\providecommand \natexlab [1]{#1}%
\providecommand \enquote  [1]{``#1''}%
\providecommand \bibnamefont  [1]{#1}%
\providecommand \bibfnamefont [1]{#1}%
\providecommand \citenamefont [1]{#1}%
\providecommand \href@noop [0]{\@secondoftwo}%
\providecommand \href [0]{\begingroup \@sanitize@url \@href}%
\providecommand \@href[1]{\@@startlink{#1}\@@href}%
\providecommand \@@href[1]{\endgroup#1\@@endlink}%
\providecommand \@sanitize@url [0]{\catcode `\\12\catcode `\$12\catcode
  `\&12\catcode `\#12\catcode `\^12\catcode `\_12\catcode `\%12\relax}%
\providecommand \@@startlink[1]{}%
\providecommand \@@endlink[0]{}%
\providecommand \url  [0]{\begingroup\@sanitize@url \@url }%
\providecommand \@url [1]{\endgroup\@href {#1}{\urlprefix }}%
\providecommand \urlprefix  [0]{URL }%
\providecommand \Eprint [0]{\href }%
\providecommand \doibase [0]{https://doi.org/}%
\providecommand \selectlanguage [0]{\@gobble}%
\providecommand \bibinfo  [0]{\@secondoftwo}%
\providecommand \bibfield  [0]{\@secondoftwo}%
\providecommand \translation [1]{[#1]}%
\providecommand \BibitemOpen [0]{}%
\providecommand \bibitemStop [0]{}%
\providecommand \bibitemNoStop [0]{.\EOS\space}%
\providecommand \EOS [0]{\spacefactor3000\relax}%
\providecommand \BibitemShut  [1]{\csname bibitem#1\endcsname}%
\let\auto@bib@innerbib\@empty
\bibitem [{\citenamefont {Hu}(2015)}]{Hu_2015}%
  \BibitemOpen
  \bibfield  {author} {\bibinfo {author} {\bibfnamefont {C.-M.}\ \bibnamefont
  {Hu}},\ }\href@noop {} {\bibfield  {journal} {\bibinfo  {journal} {arXiv
  preprint arXiv:1508.01966}\ } (\bibinfo {year} {2015})}\BibitemShut {NoStop}%
\bibitem [{\citenamefont {Harder}\ and\ \citenamefont
  {Hu}(2018)}]{Harder_2018}%
  \BibitemOpen
  \bibfield  {author} {\bibinfo {author} {\bibfnamefont {M.}~\bibnamefont
  {Harder}}\ and\ \bibinfo {author} {\bibfnamefont {C.-M.}\ \bibnamefont
  {Hu}},\ }\href@noop {} {\bibfield  {journal} {\bibinfo  {journal} {Solid
  State Physics}\ }\textbf {\bibinfo {volume} {69}},\ \bibinfo {pages} {47}
  (\bibinfo {year} {2018})}\BibitemShut {NoStop}%
\bibitem [{\citenamefont {Li}\ \emph {et~al.}(2021)\citenamefont {Li},
  \citenamefont {Zhao}, \citenamefont {Zhang}, \citenamefont {Hoffmann},\ and\
  \citenamefont {Novosad}}]{Li_Review_2021}%
  \BibitemOpen
  \bibfield  {author} {\bibinfo {author} {\bibfnamefont {Y.}~\bibnamefont
  {Li}}, \bibinfo {author} {\bibfnamefont {C.}~\bibnamefont {Zhao}}, \bibinfo
  {author} {\bibfnamefont {W.}~\bibnamefont {Zhang}}, \bibinfo {author}
  {\bibfnamefont {A.}~\bibnamefont {Hoffmann}},\ and\ \bibinfo {author}
  {\bibfnamefont {V.}~\bibnamefont {Novosad}},\ }\href@noop {} {\bibfield
  {journal} {\bibinfo  {journal} {APL Materials}\ }\textbf {\bibinfo {volume}
  {9}},\ \bibinfo {pages} {060902} (\bibinfo {year} {2021})}\BibitemShut
  {NoStop}%
\bibitem [{\citenamefont {Bhoi}\ and\ \citenamefont {Kim}(2019)}]{Bhoi_2019}%
  \BibitemOpen
  \bibfield  {author} {\bibinfo {author} {\bibfnamefont {B.}~\bibnamefont
  {Bhoi}}\ and\ \bibinfo {author} {\bibfnamefont {S.-K.}\ \bibnamefont {Kim}},\
  }in\ \href@noop {} {\emph {\bibinfo {booktitle} {Solid State Physics}}},\
  Vol.~\bibinfo {volume} {70}\ (\bibinfo  {publisher} {Elsevier},\ \bibinfo
  {year} {2019})\ pp.\ \bibinfo {pages} {1--77}\BibitemShut {NoStop}%
\bibitem [{\citenamefont {Hisatomi}\ \emph {et~al.}(2016)\citenamefont
  {Hisatomi}, \citenamefont {Osada}, \citenamefont {Tabuchi}, \citenamefont
  {Ishikawa}, \citenamefont {Noguchi}, \citenamefont {Yamazaki}, \citenamefont
  {Usami},\ and\ \citenamefont {Nakamura}}]{Hisatomi_2016}%
  \BibitemOpen
  \bibfield  {author} {\bibinfo {author} {\bibfnamefont {R.}~\bibnamefont
  {Hisatomi}}, \bibinfo {author} {\bibfnamefont {A.}~\bibnamefont {Osada}},
  \bibinfo {author} {\bibfnamefont {Y.}~\bibnamefont {Tabuchi}}, \bibinfo
  {author} {\bibfnamefont {T.}~\bibnamefont {Ishikawa}}, \bibinfo {author}
  {\bibfnamefont {A.}~\bibnamefont {Noguchi}}, \bibinfo {author} {\bibfnamefont
  {R.}~\bibnamefont {Yamazaki}}, \bibinfo {author} {\bibfnamefont
  {K.}~\bibnamefont {Usami}},\ and\ \bibinfo {author} {\bibfnamefont
  {Y.}~\bibnamefont {Nakamura}},\ }\href
  {https://doi.org/10.1103/PhysRevB.93.174427} {\bibfield  {journal} {\bibinfo
  {journal} {Phys. Rev. B}\ }\textbf {\bibinfo {volume} {93}},\ \bibinfo
  {pages} {174427} (\bibinfo {year} {2016})}\BibitemShut {NoStop}%
\bibitem [{\citenamefont {Zhang}\ \emph {et~al.}(2016)\citenamefont {Zhang},
  \citenamefont {Zhu}, \citenamefont {Zou},\ and\ \citenamefont
  {Tang}}]{XZhang_PRL_2016}%
  \BibitemOpen
  \bibfield  {author} {\bibinfo {author} {\bibfnamefont {X.}~\bibnamefont
  {Zhang}}, \bibinfo {author} {\bibfnamefont {N.}~\bibnamefont {Zhu}}, \bibinfo
  {author} {\bibfnamefont {C.-L.}\ \bibnamefont {Zou}},\ and\ \bibinfo {author}
  {\bibfnamefont {H.~X.}\ \bibnamefont {Tang}},\ }\href
  {https://doi.org/10.1103/PhysRevLett.117.123605} {\bibfield  {journal}
  {\bibinfo  {journal} {Phys. Rev. Lett.}\ }\textbf {\bibinfo {volume} {117}},\
  \bibinfo {pages} {123605} (\bibinfo {year} {2016})}\BibitemShut {NoStop}%
\bibitem [{\citenamefont {Haigh}\ \emph {et~al.}(2016)\citenamefont {Haigh},
  \citenamefont {Nunnenkamp}, \citenamefont {Ramsay},\ and\ \citenamefont
  {Ferguson}}]{Haigh_2016}%
  \BibitemOpen
  \bibfield  {author} {\bibinfo {author} {\bibfnamefont {J.~A.}\ \bibnamefont
  {Haigh}}, \bibinfo {author} {\bibfnamefont {A.}~\bibnamefont {Nunnenkamp}},
  \bibinfo {author} {\bibfnamefont {A.~J.}\ \bibnamefont {Ramsay}},\ and\
  \bibinfo {author} {\bibfnamefont {A.~J.}\ \bibnamefont {Ferguson}},\ }\href
  {https://doi.org/10.1103/PhysRevLett.117.133602} {\bibfield  {journal}
  {\bibinfo  {journal} {Phys. Rev. Lett.}\ }\textbf {\bibinfo {volume} {117}},\
  \bibinfo {pages} {133602} (\bibinfo {year} {2016})}\BibitemShut {NoStop}%
\bibitem [{\citenamefont {Osada}\ \emph {et~al.}(2018)\citenamefont {Osada},
  \citenamefont {Gloppe}, \citenamefont {Hisatomi}, \citenamefont {Noguchi},
  \citenamefont {Yamazaki}, \citenamefont {Nomura}, \citenamefont {Nakamura},\
  and\ \citenamefont {Usami}}]{Osada_2018}%
  \BibitemOpen
  \bibfield  {author} {\bibinfo {author} {\bibfnamefont {A.}~\bibnamefont
  {Osada}}, \bibinfo {author} {\bibfnamefont {A.}~\bibnamefont {Gloppe}},
  \bibinfo {author} {\bibfnamefont {R.}~\bibnamefont {Hisatomi}}, \bibinfo
  {author} {\bibfnamefont {A.}~\bibnamefont {Noguchi}}, \bibinfo {author}
  {\bibfnamefont {R.}~\bibnamefont {Yamazaki}}, \bibinfo {author}
  {\bibfnamefont {M.}~\bibnamefont {Nomura}}, \bibinfo {author} {\bibfnamefont
  {Y.}~\bibnamefont {Nakamura}},\ and\ \bibinfo {author} {\bibfnamefont
  {K.}~\bibnamefont {Usami}},\ }\href
  {https://doi.org/10.1103/PhysRevLett.120.133602} {\bibfield  {journal}
  {\bibinfo  {journal} {Phys. Rev. Lett.}\ }\textbf {\bibinfo {volume} {120}},\
  \bibinfo {pages} {133602} (\bibinfo {year} {2018})}\BibitemShut {NoStop}%
\bibitem [{\citenamefont {Klingler}\ \emph {et~al.}(2016)\citenamefont
  {Klingler}, \citenamefont {Maier-Flaig}, \citenamefont {Gross}, \citenamefont
  {Hu}, \citenamefont {Huebl}, \citenamefont {Goennenwein},\ and\ \citenamefont
  {Weiler}}]{Klingler_2016}%
  \BibitemOpen
  \bibfield  {author} {\bibinfo {author} {\bibfnamefont {S.}~\bibnamefont
  {Klingler}}, \bibinfo {author} {\bibfnamefont {H.}~\bibnamefont
  {Maier-Flaig}}, \bibinfo {author} {\bibfnamefont {R.}~\bibnamefont {Gross}},
  \bibinfo {author} {\bibfnamefont {C.~M.}\ \bibnamefont {Hu}}, \bibinfo
  {author} {\bibfnamefont {H.}~\bibnamefont {Huebl}}, \bibinfo {author}
  {\bibfnamefont {S.~T.~B.}\ \bibnamefont {Goennenwein}},\ and\ \bibinfo
  {author} {\bibfnamefont {M.}~\bibnamefont {Weiler}},\ }\href
  {https://doi.org/10.1063/1.4961052} {\bibfield  {journal} {\bibinfo
  {journal} {Appl. Phys. Lett.}\ }\textbf {\bibinfo {volume} {109}},\ \bibinfo
  {pages} {072402} (\bibinfo {year} {2016})}\BibitemShut {NoStop}%
\bibitem [{\citenamefont {Schultheiss}\ \emph {et~al.}(2019)\citenamefont
  {Schultheiss}, \citenamefont {Verba}, \citenamefont {Wehrmann}, \citenamefont
  {Wagner}, \citenamefont {K\"orber}, \citenamefont {Hula}, \citenamefont
  {Hache}, \citenamefont {K\'akay}, \citenamefont {Awad}, \citenamefont
  {Tiberkevich}, \citenamefont {Slavin}, \citenamefont {Fassbender},\ and\
  \citenamefont {Schultheiss}}]{Schultheiss_2019}%
  \BibitemOpen
  \bibfield  {author} {\bibinfo {author} {\bibfnamefont {K.}~\bibnamefont
  {Schultheiss}}, \bibinfo {author} {\bibfnamefont {R.}~\bibnamefont {Verba}},
  \bibinfo {author} {\bibfnamefont {F.}~\bibnamefont {Wehrmann}}, \bibinfo
  {author} {\bibfnamefont {K.}~\bibnamefont {Wagner}}, \bibinfo {author}
  {\bibfnamefont {L.}~\bibnamefont {K\"orber}}, \bibinfo {author}
  {\bibfnamefont {T.}~\bibnamefont {Hula}}, \bibinfo {author} {\bibfnamefont
  {T.}~\bibnamefont {Hache}}, \bibinfo {author} {\bibfnamefont
  {A.}~\bibnamefont {K\'akay}}, \bibinfo {author} {\bibfnamefont {A.~A.}\
  \bibnamefont {Awad}}, \bibinfo {author} {\bibfnamefont {V.}~\bibnamefont
  {Tiberkevich}}, \bibinfo {author} {\bibfnamefont {A.~N.}\ \bibnamefont
  {Slavin}}, \bibinfo {author} {\bibfnamefont {J.}~\bibnamefont {Fassbender}},\
  and\ \bibinfo {author} {\bibfnamefont {H.}~\bibnamefont {Schultheiss}},\
  }\href {https://doi.org/10.1103/PhysRevLett.122.097202} {\bibfield  {journal}
  {\bibinfo  {journal} {Phys. Rev. Lett.}\ }\textbf {\bibinfo {volume} {122}},\
  \bibinfo {pages} {097202} (\bibinfo {year} {2019})}\BibitemShut {NoStop}%
\bibitem [{\citenamefont {Tabuchi}\ \emph {et~al.}(2014)\citenamefont
  {Tabuchi}, \citenamefont {Ishino}, \citenamefont {Ishikawa}, \citenamefont
  {Yamazaki}, \citenamefont {Usami},\ and\ \citenamefont
  {Nakamura}}]{Tabuchi_2014}%
  \BibitemOpen
  \bibfield  {author} {\bibinfo {author} {\bibfnamefont {Y.}~\bibnamefont
  {Tabuchi}}, \bibinfo {author} {\bibfnamefont {S.}~\bibnamefont {Ishino}},
  \bibinfo {author} {\bibfnamefont {T.}~\bibnamefont {Ishikawa}}, \bibinfo
  {author} {\bibfnamefont {R.}~\bibnamefont {Yamazaki}}, \bibinfo {author}
  {\bibfnamefont {K.}~\bibnamefont {Usami}},\ and\ \bibinfo {author}
  {\bibfnamefont {Y.}~\bibnamefont {Nakamura}},\ }\href@noop {} {\bibfield
  {journal} {\bibinfo  {journal} {Physical review letters}\ }\textbf {\bibinfo
  {volume} {113}},\ \bibinfo {pages} {083603} (\bibinfo {year}
  {2014})}\BibitemShut {NoStop}%
\bibitem [{\citenamefont {Bhoi}\ \emph {et~al.}(2014)\citenamefont {Bhoi},
  \citenamefont {Cliff}, \citenamefont {Maksymov}, \citenamefont {Kostylev},
  \citenamefont {Aiyar}, \citenamefont {Venkataramani}, \citenamefont
  {Prasad},\ and\ \citenamefont {Stamps}}]{Bhoi_2014}%
  \BibitemOpen
  \bibfield  {author} {\bibinfo {author} {\bibfnamefont {B.}~\bibnamefont
  {Bhoi}}, \bibinfo {author} {\bibfnamefont {T.}~\bibnamefont {Cliff}},
  \bibinfo {author} {\bibfnamefont {I.}~\bibnamefont {Maksymov}}, \bibinfo
  {author} {\bibfnamefont {M.}~\bibnamefont {Kostylev}}, \bibinfo {author}
  {\bibfnamefont {R.}~\bibnamefont {Aiyar}}, \bibinfo {author} {\bibfnamefont
  {N.}~\bibnamefont {Venkataramani}}, \bibinfo {author} {\bibfnamefont
  {S.}~\bibnamefont {Prasad}},\ and\ \bibinfo {author} {\bibfnamefont
  {R.}~\bibnamefont {Stamps}},\ }\href@noop {} {\bibfield  {journal} {\bibinfo
  {journal} {Journal of Applied Physics}\ }\textbf {\bibinfo {volume} {116}},\
  \bibinfo {pages} {243906} (\bibinfo {year} {2014})}\BibitemShut {NoStop}%
\bibitem [{\citenamefont {Harder}\ \emph {et~al.}(2018)\citenamefont {Harder},
  \citenamefont {Yang}, \citenamefont {Yao}, \citenamefont {Yu}, \citenamefont
  {Rao}, \citenamefont {Gui}, \citenamefont {Stamps},\ and\ \citenamefont
  {Hu}}]{Harder_PRL2018}%
  \BibitemOpen
  \bibfield  {author} {\bibinfo {author} {\bibfnamefont {M.}~\bibnamefont
  {Harder}}, \bibinfo {author} {\bibfnamefont {Y.}~\bibnamefont {Yang}},
  \bibinfo {author} {\bibfnamefont {B.~M.}\ \bibnamefont {Yao}}, \bibinfo
  {author} {\bibfnamefont {C.~H.}\ \bibnamefont {Yu}}, \bibinfo {author}
  {\bibfnamefont {J.~W.}\ \bibnamefont {Rao}}, \bibinfo {author} {\bibfnamefont
  {Y.~S.}\ \bibnamefont {Gui}}, \bibinfo {author} {\bibfnamefont {R.~L.}\
  \bibnamefont {Stamps}},\ and\ \bibinfo {author} {\bibfnamefont {C.-M.}\
  \bibnamefont {Hu}},\ }\href {https://doi.org/10.1103/PhysRevLett.121.137203}
  {\bibfield  {journal} {\bibinfo  {journal} {Phys. Rev. Lett.}\ }\textbf
  {\bibinfo {volume} {121}},\ \bibinfo {pages} {137203} (\bibinfo {year}
  {2018})}\BibitemShut {NoStop}%
\bibitem [{\citenamefont {Bai}\ \emph {et~al.}(2015)\citenamefont {Bai},
  \citenamefont {Harder}, \citenamefont {Chen}, \citenamefont {Fan},
  \citenamefont {Xiao},\ and\ \citenamefont {Hu}}]{Bai_2015}%
  \BibitemOpen
  \bibfield  {author} {\bibinfo {author} {\bibfnamefont {L.}~\bibnamefont
  {Bai}}, \bibinfo {author} {\bibfnamefont {M.}~\bibnamefont {Harder}},
  \bibinfo {author} {\bibfnamefont {Y.~P.}\ \bibnamefont {Chen}}, \bibinfo
  {author} {\bibfnamefont {X.}~\bibnamefont {Fan}}, \bibinfo {author}
  {\bibfnamefont {J.~Q.}\ \bibnamefont {Xiao}},\ and\ \bibinfo {author}
  {\bibfnamefont {C.-M.}\ \bibnamefont {Hu}},\ }\href
  {https://doi.org/10.1103/PhysRevLett.114.227201} {\bibfield  {journal}
  {\bibinfo  {journal} {Phys. Rev. Lett.}\ }\textbf {\bibinfo {volume} {114}},\
  \bibinfo {pages} {227201} (\bibinfo {year} {2015})}\BibitemShut {NoStop}%
\bibitem [{\citenamefont {Castel}\ \emph {et~al.}(2017)\citenamefont {Castel},
  \citenamefont {Manchec},\ and\ \citenamefont {Ben~Youssef}}]{Castel2017}%
  \BibitemOpen
  \bibfield  {author} {\bibinfo {author} {\bibfnamefont {V.}~\bibnamefont
  {Castel}}, \bibinfo {author} {\bibfnamefont {A.}~\bibnamefont {Manchec}},\
  and\ \bibinfo {author} {\bibfnamefont {J.}~\bibnamefont {Ben~Youssef}},\
  }\href {https://doi.org/10.1109/LMAG.2016.2635627} {\bibfield  {journal}
  {\bibinfo  {journal} {IEEE Magnetics Letters}\ }\textbf {\bibinfo {volume}
  {8}},\ \bibinfo {pages} {1} (\bibinfo {year} {2017})}\BibitemShut {NoStop}%
\bibitem [{\citenamefont {Chumak}\ \emph {et~al.}(2021)\citenamefont {Chumak},
  \citenamefont {Kabos}, \citenamefont {Wu}, \citenamefont {Abert},
  \citenamefont {Adelmann}, \citenamefont {Adeyeye}, \citenamefont {Åkerman},
  \citenamefont {Aliev}, \citenamefont {Anane}, \citenamefont {Awad},
  \citenamefont {Back}, \citenamefont {Barman}, \citenamefont {Bauer},
  \citenamefont {Becherer}, \citenamefont {Beginin}, \citenamefont
  {Bittencourt}, \citenamefont {Blanter}, \citenamefont {Bortolotti},
  \citenamefont {Boventer}, \citenamefont {Bozhko}, \citenamefont {Bunyaev},
  \citenamefont {Carmiggelt}, \citenamefont {Cheenikundil}, \citenamefont
  {Ciubotaru}, \citenamefont {Cotofana}, \citenamefont {Csaba}, \citenamefont
  {Dobrovolskiy}, \citenamefont {Dubs}, \citenamefont {Elyasi}, \citenamefont
  {Fripp}, \citenamefont {Fulara}, \citenamefont {Golovchanskiy}, \citenamefont
  {Gonzalez-Ballestero}, \citenamefont {Graczyk}, \citenamefont {Grundler},
  \citenamefont {Gruszecki}, \citenamefont {Gubbiotti}, \citenamefont
  {Guslienko}, \citenamefont {Haldar}, \citenamefont {Hamdioui}, \citenamefont
  {Hertel}, \citenamefont {Hillebrands}, \citenamefont {Hioki}, \citenamefont
  {Houshang}, \citenamefont {Hu}, \citenamefont {Huebl}, \citenamefont {Huth},
  \citenamefont {Iacocca}, \citenamefont {Jungfleisch}, \citenamefont
  {Kakazei}, \citenamefont {Khitun}, \citenamefont {Khymyn}, \citenamefont
  {Kikkawa}, \citenamefont {Kläui}, \citenamefont {Klein}, \citenamefont
  {Kłos}, \citenamefont {Knauer}, \citenamefont {Koraltan}, \citenamefont
  {Kostylev}, \citenamefont {Krawczyk}, \citenamefont {Krivorotov},
  \citenamefont {Kruglyak}, \citenamefont {Lachance-Quirion}, \citenamefont
  {Ladak}, \citenamefont {Lebrun}, \citenamefont {Li}, \citenamefont {Lindner},
  \citenamefont {Macêdo}, \citenamefont {Mayr}, \citenamefont {Melkov},
  \citenamefont {Mieszczak}, \citenamefont {Nakamura}, \citenamefont {Nembach},
  \citenamefont {Nikitin}, \citenamefont {Nikitov}, \citenamefont {Novosad},
  \citenamefont {Otalora}, \citenamefont {Otani}, \citenamefont {Papp},
  \citenamefont {Pigeau}, \citenamefont {Pirro}, \citenamefont {Porod},
  \citenamefont {Porrati}, \citenamefont {Qin}, \citenamefont {Rana},
  \citenamefont {Reimann}, \citenamefont {Riente}, \citenamefont
  {Romero-Isart}, \citenamefont {Ross}, \citenamefont {Sadovnikov},
  \citenamefont {Safin}, \citenamefont {Saitoh}, \citenamefont {Schmidt},
  \citenamefont {Schultheiss}, \citenamefont {Schultheiss}, \citenamefont
  {Serga}, \citenamefont {Sharma}, \citenamefont {Shaw}, \citenamefont {Suess},
  \citenamefont {Surzhenko}, \citenamefont {Szulc}, \citenamefont {Taniguchi},
  \citenamefont {Urbánek}, \citenamefont {Usami}, \citenamefont {Ustinov},
  \citenamefont {van~der Sar}, \citenamefont {van Dijken}, \citenamefont
  {Vasyuchka}, \citenamefont {Verba}, \citenamefont {Kusminskiy}, \citenamefont
  {Wang}, \citenamefont {Weides}, \citenamefont {Weiler}, \citenamefont
  {Wintz}, \citenamefont {Wolski},\ and\ \citenamefont
  {Zhang}}]{chumak2021roadmap}%
  \BibitemOpen
  \bibfield  {author} {\bibinfo {author} {\bibfnamefont {A.~V.}\ \bibnamefont
  {Chumak}}, \bibinfo {author} {\bibfnamefont {P.}~\bibnamefont {Kabos}},
  \bibinfo {author} {\bibfnamefont {M.}~\bibnamefont {Wu}}, \bibinfo {author}
  {\bibfnamefont {C.}~\bibnamefont {Abert}}, \bibinfo {author} {\bibfnamefont
  {C.}~\bibnamefont {Adelmann}}, \bibinfo {author} {\bibfnamefont
  {A.}~\bibnamefont {Adeyeye}}, \bibinfo {author} {\bibfnamefont
  {J.}~\bibnamefont {Åkerman}}, \bibinfo {author} {\bibfnamefont {F.~G.}\
  \bibnamefont {Aliev}}, \bibinfo {author} {\bibfnamefont {A.}~\bibnamefont
  {Anane}}, \bibinfo {author} {\bibfnamefont {A.}~\bibnamefont {Awad}},
  \bibinfo {author} {\bibfnamefont {C.~H.}\ \bibnamefont {Back}}, \bibinfo
  {author} {\bibfnamefont {A.}~\bibnamefont {Barman}}, \bibinfo {author}
  {\bibfnamefont {G.~E.~W.}\ \bibnamefont {Bauer}}, \bibinfo {author}
  {\bibfnamefont {M.}~\bibnamefont {Becherer}}, \bibinfo {author}
  {\bibfnamefont {E.~N.}\ \bibnamefont {Beginin}}, \bibinfo {author}
  {\bibfnamefont {V.~A. S.~V.}\ \bibnamefont {Bittencourt}}, \bibinfo {author}
  {\bibfnamefont {Y.~M.}\ \bibnamefont {Blanter}}, \bibinfo {author}
  {\bibfnamefont {P.}~\bibnamefont {Bortolotti}}, \bibinfo {author}
  {\bibfnamefont {I.}~\bibnamefont {Boventer}}, \bibinfo {author}
  {\bibfnamefont {D.~A.}\ \bibnamefont {Bozhko}}, \bibinfo {author}
  {\bibfnamefont {S.~A.}\ \bibnamefont {Bunyaev}}, \bibinfo {author}
  {\bibfnamefont {J.~J.}\ \bibnamefont {Carmiggelt}}, \bibinfo {author}
  {\bibfnamefont {R.~R.}\ \bibnamefont {Cheenikundil}}, \bibinfo {author}
  {\bibfnamefont {F.}~\bibnamefont {Ciubotaru}}, \bibinfo {author}
  {\bibfnamefont {S.}~\bibnamefont {Cotofana}}, \bibinfo {author}
  {\bibfnamefont {G.}~\bibnamefont {Csaba}}, \bibinfo {author} {\bibfnamefont
  {O.~V.}\ \bibnamefont {Dobrovolskiy}}, \bibinfo {author} {\bibfnamefont
  {C.}~\bibnamefont {Dubs}}, \bibinfo {author} {\bibfnamefont {M.}~\bibnamefont
  {Elyasi}}, \bibinfo {author} {\bibfnamefont {K.~G.}\ \bibnamefont {Fripp}},
  \bibinfo {author} {\bibfnamefont {H.}~\bibnamefont {Fulara}}, \bibinfo
  {author} {\bibfnamefont {I.~A.}\ \bibnamefont {Golovchanskiy}}, \bibinfo
  {author} {\bibfnamefont {C.}~\bibnamefont {Gonzalez-Ballestero}}, \bibinfo
  {author} {\bibfnamefont {P.}~\bibnamefont {Graczyk}}, \bibinfo {author}
  {\bibfnamefont {D.}~\bibnamefont {Grundler}}, \bibinfo {author}
  {\bibfnamefont {P.}~\bibnamefont {Gruszecki}}, \bibinfo {author}
  {\bibfnamefont {G.}~\bibnamefont {Gubbiotti}}, \bibinfo {author}
  {\bibfnamefont {K.}~\bibnamefont {Guslienko}}, \bibinfo {author}
  {\bibfnamefont {A.}~\bibnamefont {Haldar}}, \bibinfo {author} {\bibfnamefont
  {S.}~\bibnamefont {Hamdioui}}, \bibinfo {author} {\bibfnamefont
  {R.}~\bibnamefont {Hertel}}, \bibinfo {author} {\bibfnamefont
  {B.}~\bibnamefont {Hillebrands}}, \bibinfo {author} {\bibfnamefont
  {T.}~\bibnamefont {Hioki}}, \bibinfo {author} {\bibfnamefont
  {A.}~\bibnamefont {Houshang}}, \bibinfo {author} {\bibfnamefont {C.~M.}\
  \bibnamefont {Hu}}, \bibinfo {author} {\bibfnamefont {H.}~\bibnamefont
  {Huebl}}, \bibinfo {author} {\bibfnamefont {M.}~\bibnamefont {Huth}},
  \bibinfo {author} {\bibfnamefont {E.}~\bibnamefont {Iacocca}}, \bibinfo
  {author} {\bibfnamefont {M.~B.}\ \bibnamefont {Jungfleisch}}, \bibinfo
  {author} {\bibfnamefont {G.~N.}\ \bibnamefont {Kakazei}}, \bibinfo {author}
  {\bibfnamefont {A.}~\bibnamefont {Khitun}}, \bibinfo {author} {\bibfnamefont
  {R.}~\bibnamefont {Khymyn}}, \bibinfo {author} {\bibfnamefont
  {T.}~\bibnamefont {Kikkawa}}, \bibinfo {author} {\bibfnamefont
  {M.}~\bibnamefont {Kläui}}, \bibinfo {author} {\bibfnamefont
  {O.}~\bibnamefont {Klein}}, \bibinfo {author} {\bibfnamefont {J.~W.}\
  \bibnamefont {Kłos}}, \bibinfo {author} {\bibfnamefont {S.}~\bibnamefont
  {Knauer}}, \bibinfo {author} {\bibfnamefont {S.}~\bibnamefont {Koraltan}},
  \bibinfo {author} {\bibfnamefont {M.}~\bibnamefont {Kostylev}}, \bibinfo
  {author} {\bibfnamefont {M.}~\bibnamefont {Krawczyk}}, \bibinfo {author}
  {\bibfnamefont {I.~N.}\ \bibnamefont {Krivorotov}}, \bibinfo {author}
  {\bibfnamefont {V.~V.}\ \bibnamefont {Kruglyak}}, \bibinfo {author}
  {\bibfnamefont {D.}~\bibnamefont {Lachance-Quirion}}, \bibinfo {author}
  {\bibfnamefont {S.}~\bibnamefont {Ladak}}, \bibinfo {author} {\bibfnamefont
  {R.}~\bibnamefont {Lebrun}}, \bibinfo {author} {\bibfnamefont
  {Y.}~\bibnamefont {Li}}, \bibinfo {author} {\bibfnamefont {M.}~\bibnamefont
  {Lindner}}, \bibinfo {author} {\bibfnamefont {R.}~\bibnamefont {Macêdo}},
  \bibinfo {author} {\bibfnamefont {S.}~\bibnamefont {Mayr}}, \bibinfo {author}
  {\bibfnamefont {G.~A.}\ \bibnamefont {Melkov}}, \bibinfo {author}
  {\bibfnamefont {S.}~\bibnamefont {Mieszczak}}, \bibinfo {author}
  {\bibfnamefont {Y.}~\bibnamefont {Nakamura}}, \bibinfo {author}
  {\bibfnamefont {H.~T.}\ \bibnamefont {Nembach}}, \bibinfo {author}
  {\bibfnamefont {A.~A.}\ \bibnamefont {Nikitin}}, \bibinfo {author}
  {\bibfnamefont {S.~A.}\ \bibnamefont {Nikitov}}, \bibinfo {author}
  {\bibfnamefont {V.}~\bibnamefont {Novosad}}, \bibinfo {author} {\bibfnamefont
  {J.~A.}\ \bibnamefont {Otalora}}, \bibinfo {author} {\bibfnamefont
  {Y.}~\bibnamefont {Otani}}, \bibinfo {author} {\bibfnamefont
  {A.}~\bibnamefont {Papp}}, \bibinfo {author} {\bibfnamefont {B.}~\bibnamefont
  {Pigeau}}, \bibinfo {author} {\bibfnamefont {P.}~\bibnamefont {Pirro}},
  \bibinfo {author} {\bibfnamefont {W.}~\bibnamefont {Porod}}, \bibinfo
  {author} {\bibfnamefont {F.}~\bibnamefont {Porrati}}, \bibinfo {author}
  {\bibfnamefont {H.}~\bibnamefont {Qin}}, \bibinfo {author} {\bibfnamefont
  {B.}~\bibnamefont {Rana}}, \bibinfo {author} {\bibfnamefont {T.}~\bibnamefont
  {Reimann}}, \bibinfo {author} {\bibfnamefont {F.}~\bibnamefont {Riente}},
  \bibinfo {author} {\bibfnamefont {O.}~\bibnamefont {Romero-Isart}}, \bibinfo
  {author} {\bibfnamefont {A.}~\bibnamefont {Ross}}, \bibinfo {author}
  {\bibfnamefont {A.~V.}\ \bibnamefont {Sadovnikov}}, \bibinfo {author}
  {\bibfnamefont {A.~R.}\ \bibnamefont {Safin}}, \bibinfo {author}
  {\bibfnamefont {E.}~\bibnamefont {Saitoh}}, \bibinfo {author} {\bibfnamefont
  {G.}~\bibnamefont {Schmidt}}, \bibinfo {author} {\bibfnamefont
  {H.}~\bibnamefont {Schultheiss}}, \bibinfo {author} {\bibfnamefont
  {K.}~\bibnamefont {Schultheiss}}, \bibinfo {author} {\bibfnamefont {A.~A.}\
  \bibnamefont {Serga}}, \bibinfo {author} {\bibfnamefont {S.}~\bibnamefont
  {Sharma}}, \bibinfo {author} {\bibfnamefont {J.~M.}\ \bibnamefont {Shaw}},
  \bibinfo {author} {\bibfnamefont {D.}~\bibnamefont {Suess}}, \bibinfo
  {author} {\bibfnamefont {O.}~\bibnamefont {Surzhenko}}, \bibinfo {author}
  {\bibfnamefont {K.}~\bibnamefont {Szulc}}, \bibinfo {author} {\bibfnamefont
  {T.}~\bibnamefont {Taniguchi}}, \bibinfo {author} {\bibfnamefont
  {M.}~\bibnamefont {Urbánek}}, \bibinfo {author} {\bibfnamefont
  {K.}~\bibnamefont {Usami}}, \bibinfo {author} {\bibfnamefont {A.~B.}\
  \bibnamefont {Ustinov}}, \bibinfo {author} {\bibfnamefont {T.}~\bibnamefont
  {van~der Sar}}, \bibinfo {author} {\bibfnamefont {S.}~\bibnamefont {van
  Dijken}}, \bibinfo {author} {\bibfnamefont {V.~I.}\ \bibnamefont
  {Vasyuchka}}, \bibinfo {author} {\bibfnamefont {R.}~\bibnamefont {Verba}},
  \bibinfo {author} {\bibfnamefont {S.~V.}\ \bibnamefont {Kusminskiy}},
  \bibinfo {author} {\bibfnamefont {Q.}~\bibnamefont {Wang}}, \bibinfo {author}
  {\bibfnamefont {M.}~\bibnamefont {Weides}}, \bibinfo {author} {\bibfnamefont
  {M.}~\bibnamefont {Weiler}}, \bibinfo {author} {\bibfnamefont
  {S.}~\bibnamefont {Wintz}}, \bibinfo {author} {\bibfnamefont {S.~P.}\
  \bibnamefont {Wolski}},\ and\ \bibinfo {author} {\bibfnamefont
  {X.}~\bibnamefont {Zhang}},\ }\href@noop {} {\bibinfo {title} {Roadmap on
  spin-wave computing}} (\bibinfo {year} {2021}),\ \Eprint
  {https://arxiv.org/abs/2111.00365} {arXiv:2111.00365 [physics.app-ph]}
  \BibitemShut {NoStop}%
\bibitem [{\citenamefont {Li}\ \emph {et~al.}(2019)\citenamefont {Li},
  \citenamefont {Polakovic}, \citenamefont {Wang}, \citenamefont {Xu},
  \citenamefont {Lendinez}, \citenamefont {Zhang}, \citenamefont {Ding},
  \citenamefont {Khaire}, \citenamefont {Saglam}, \citenamefont {Divan},
  \citenamefont {Pearson}, \citenamefont {Kwok}, \citenamefont {Xiao},
  \citenamefont {Novosad}, \citenamefont {Hoffmann},\ and\ \citenamefont
  {Zhang}}]{Li_PRL2019}%
  \BibitemOpen
  \bibfield  {author} {\bibinfo {author} {\bibfnamefont {Y.}~\bibnamefont
  {Li}}, \bibinfo {author} {\bibfnamefont {T.}~\bibnamefont {Polakovic}},
  \bibinfo {author} {\bibfnamefont {Y.-L.}\ \bibnamefont {Wang}}, \bibinfo
  {author} {\bibfnamefont {J.}~\bibnamefont {Xu}}, \bibinfo {author}
  {\bibfnamefont {S.}~\bibnamefont {Lendinez}}, \bibinfo {author}
  {\bibfnamefont {Z.}~\bibnamefont {Zhang}}, \bibinfo {author} {\bibfnamefont
  {J.}~\bibnamefont {Ding}}, \bibinfo {author} {\bibfnamefont {T.}~\bibnamefont
  {Khaire}}, \bibinfo {author} {\bibfnamefont {H.}~\bibnamefont {Saglam}},
  \bibinfo {author} {\bibfnamefont {R.}~\bibnamefont {Divan}}, \bibinfo
  {author} {\bibfnamefont {J.}~\bibnamefont {Pearson}}, \bibinfo {author}
  {\bibfnamefont {W.-K.}\ \bibnamefont {Kwok}}, \bibinfo {author}
  {\bibfnamefont {Z.}~\bibnamefont {Xiao}}, \bibinfo {author} {\bibfnamefont
  {V.}~\bibnamefont {Novosad}}, \bibinfo {author} {\bibfnamefont
  {A.}~\bibnamefont {Hoffmann}},\ and\ \bibinfo {author} {\bibfnamefont
  {W.}~\bibnamefont {Zhang}},\ }\href
  {https://doi.org/10.1103/PhysRevLett.123.107701} {\bibfield  {journal}
  {\bibinfo  {journal} {Phys. Rev. Lett.}\ }\textbf {\bibinfo {volume} {123}},\
  \bibinfo {pages} {107701} (\bibinfo {year} {2019})}\BibitemShut {NoStop}%
\bibitem [{\citenamefont {Hou}\ and\ \citenamefont {Liu}(2019)}]{Hou_2019}%
  \BibitemOpen
  \bibfield  {author} {\bibinfo {author} {\bibfnamefont {J.~T.}\ \bibnamefont
  {Hou}}\ and\ \bibinfo {author} {\bibfnamefont {L.}~\bibnamefont {Liu}},\
  }\href {https://doi.org/10.1103/PhysRevLett.123.107702} {\bibfield  {journal}
  {\bibinfo  {journal} {Phys. Rev. Lett.}\ }\textbf {\bibinfo {volume} {123}},\
  \bibinfo {pages} {107702} (\bibinfo {year} {2019})}\BibitemShut {NoStop}%
\bibitem [{\citenamefont {Kargar}\ and\ \citenamefont
  {Balandin}(2021)}]{Kargar_2021}%
  \BibitemOpen
  \bibfield  {author} {\bibinfo {author} {\bibfnamefont {F.}~\bibnamefont
  {Kargar}}\ and\ \bibinfo {author} {\bibfnamefont {A.~A.}\ \bibnamefont
  {Balandin}},\ }\href {https://doi.org/10.1038/s41566-021-00836-5} {\bibfield
  {journal} {\bibinfo  {journal} {Nat. Photon.}\ ,\ \bibinfo {pages} {1}}
  (\bibinfo {year} {2021})}\BibitemShut {NoStop}%
\bibitem [{\citenamefont {Jungfleisch}(2020)}]{Jungfleisch_2020}%
  \BibitemOpen
  \bibfield  {author} {\bibinfo {author} {\bibfnamefont {M.~B.}\ \bibnamefont
  {Jungfleisch}},\ }\href@noop {} {\emph {\bibinfo {title} {{Inelastic
  Scattering of Light by Spin Waves. Optomagnonic Structures}}}},\ edited by\
  \bibinfo {editor} {\bibfnamefont {E.}~\bibnamefont {Almpanis}}\ (\bibinfo
  {publisher} {World Scientific},\ \bibinfo {year} {2020})\BibitemShut
  {NoStop}%
\bibitem [{\citenamefont {Madami}\ \emph {et~al.}(2012)\citenamefont {Madami},
  \citenamefont {Gubbiotti}, \citenamefont {Tacchi},\ and\ \citenamefont
  {Carlotti}}]{madami_2012_book}%
  \BibitemOpen
  \bibfield  {author} {\bibinfo {author} {\bibfnamefont {M.}~\bibnamefont
  {Madami}}, \bibinfo {author} {\bibfnamefont {G.}~\bibnamefont {Gubbiotti}},
  \bibinfo {author} {\bibfnamefont {S.}~\bibnamefont {Tacchi}},\ and\ \bibinfo
  {author} {\bibfnamefont {G.}~\bibnamefont {Carlotti}},\ }in\ \href@noop {}
  {\emph {\bibinfo {booktitle} {Solid state physics}}},\ Vol.~\bibinfo {volume}
  {63}\ (\bibinfo  {publisher} {Elsevier},\ \bibinfo {year} {2012})\ pp.\
  \bibinfo {pages} {79--150}\BibitemShut {NoStop}%
\bibitem [{\citenamefont {Zhang}\ \emph {et~al.}(2014)\citenamefont {Zhang},
  \citenamefont {Zou}, \citenamefont {Jiang},\ and\ \citenamefont
  {Tang}}]{zhang_2014_PRL}%
  \BibitemOpen
  \bibfield  {author} {\bibinfo {author} {\bibfnamefont {X.}~\bibnamefont
  {Zhang}}, \bibinfo {author} {\bibfnamefont {C.-L.}\ \bibnamefont {Zou}},
  \bibinfo {author} {\bibfnamefont {L.}~\bibnamefont {Jiang}},\ and\ \bibinfo
  {author} {\bibfnamefont {H.~X.}\ \bibnamefont {Tang}},\ }\href@noop {}
  {\bibfield  {journal} {\bibinfo  {journal} {Physical review letters}\
  }\textbf {\bibinfo {volume} {113}},\ \bibinfo {pages} {156401} (\bibinfo
  {year} {2014})}\BibitemShut {NoStop}%
\bibitem [{\citenamefont {Rameshti}\ \emph {et~al.}(2021)\citenamefont
  {Rameshti}, \citenamefont {Kusminskiy}, \citenamefont {Haigh}, \citenamefont
  {Usami}, \citenamefont {Lachance-Quirion}, \citenamefont {Nakamura},
  \citenamefont {Hu}, \citenamefont {Tang}, \citenamefont {Bauer},\ and\
  \citenamefont {Blanter}}]{rameshti_2021_arXiv}%
  \BibitemOpen
  \bibfield  {author} {\bibinfo {author} {\bibfnamefont {B.~Z.}\ \bibnamefont
  {Rameshti}}, \bibinfo {author} {\bibfnamefont {S.~V.}\ \bibnamefont
  {Kusminskiy}}, \bibinfo {author} {\bibfnamefont {J.~A.}\ \bibnamefont
  {Haigh}}, \bibinfo {author} {\bibfnamefont {K.}~\bibnamefont {Usami}},
  \bibinfo {author} {\bibfnamefont {D.}~\bibnamefont {Lachance-Quirion}},
  \bibinfo {author} {\bibfnamefont {Y.}~\bibnamefont {Nakamura}}, \bibinfo
  {author} {\bibfnamefont {C.-M.}\ \bibnamefont {Hu}}, \bibinfo {author}
  {\bibfnamefont {H.~X.}\ \bibnamefont {Tang}}, \bibinfo {author}
  {\bibfnamefont {G.~E.}\ \bibnamefont {Bauer}},\ and\ \bibinfo {author}
  {\bibfnamefont {Y.~M.}\ \bibnamefont {Blanter}},\ }\href@noop {} {\bibfield
  {journal} {\bibinfo  {journal} {arXiv preprint arXiv:2106.09312}\ } (\bibinfo
  {year} {2021})}\BibitemShut {NoStop}%
\bibitem [{\citenamefont {Bhoi}\ and\ \citenamefont
  {Kim}(2020)}]{Bhoi_2020_book}%
  \BibitemOpen
  \bibfield  {author} {\bibinfo {author} {\bibfnamefont {B.}~\bibnamefont
  {Bhoi}}\ and\ \bibinfo {author} {\bibfnamefont {S.-K.}\ \bibnamefont {Kim}},\
  }in\ \href@noop {} {\emph {\bibinfo {booktitle} {Solid State Physics}}},\
  Vol.~\bibinfo {volume} {71}\ (\bibinfo  {publisher} {Elsevier},\ \bibinfo
  {year} {2020})\ pp.\ \bibinfo {pages} {39--71}\BibitemShut {NoStop}%
\bibitem [{\citenamefont {Stenning}\ \emph {et~al.}(2013)\citenamefont
  {Stenning}, \citenamefont {Bowden}, \citenamefont {Maple}, \citenamefont
  {Gregory}, \citenamefont {Sposito}, \citenamefont {Eason}, \citenamefont
  {Zheludev},\ and\ \citenamefont {de~Groot}}]{stenning_2013_Optica}%
  \BibitemOpen
  \bibfield  {author} {\bibinfo {author} {\bibfnamefont {G.~B.}\ \bibnamefont
  {Stenning}}, \bibinfo {author} {\bibfnamefont {G.~J.}\ \bibnamefont
  {Bowden}}, \bibinfo {author} {\bibfnamefont {L.~C.}\ \bibnamefont {Maple}},
  \bibinfo {author} {\bibfnamefont {S.~A.}\ \bibnamefont {Gregory}}, \bibinfo
  {author} {\bibfnamefont {A.}~\bibnamefont {Sposito}}, \bibinfo {author}
  {\bibfnamefont {R.~W.}\ \bibnamefont {Eason}}, \bibinfo {author}
  {\bibfnamefont {N.~I.}\ \bibnamefont {Zheludev}},\ and\ \bibinfo {author}
  {\bibfnamefont {P.~A.}\ \bibnamefont {de~Groot}},\ }\href@noop {} {\bibfield
  {journal} {\bibinfo  {journal} {Optics Express}\ }\textbf {\bibinfo {volume}
  {21}},\ \bibinfo {pages} {1456} (\bibinfo {year} {2013})}\BibitemShut
  {NoStop}%
\bibitem [{\citenamefont {Zhang}\ \emph {et~al.}(2017)\citenamefont {Zhang},
  \citenamefont {Song},\ and\ \citenamefont {Chai}}]{DZhang_2017_JPD}%
  \BibitemOpen
  \bibfield  {author} {\bibinfo {author} {\bibfnamefont {D.}~\bibnamefont
  {Zhang}}, \bibinfo {author} {\bibfnamefont {W.}~\bibnamefont {Song}},\ and\
  \bibinfo {author} {\bibfnamefont {G.}~\bibnamefont {Chai}},\ }\href@noop {}
  {\bibfield  {journal} {\bibinfo  {journal} {Journal of Physics D: Applied
  Physics}\ }\textbf {\bibinfo {volume} {50}},\ \bibinfo {pages} {205003}
  (\bibinfo {year} {2017})}\BibitemShut {NoStop}%
\bibitem [{\citenamefont {Lambert}\ \emph {et~al.}(2015)\citenamefont
  {Lambert}, \citenamefont {Haigh},\ and\ \citenamefont
  {Ferguson}}]{lambert_2015_JAP}%
  \BibitemOpen
  \bibfield  {author} {\bibinfo {author} {\bibfnamefont {N.}~\bibnamefont
  {Lambert}}, \bibinfo {author} {\bibfnamefont {J.}~\bibnamefont {Haigh}},\
  and\ \bibinfo {author} {\bibfnamefont {A.}~\bibnamefont {Ferguson}},\
  }\href@noop {} {\bibfield  {journal} {\bibinfo  {journal} {Journal of Applied
  Physics}\ }\textbf {\bibinfo {volume} {117}},\ \bibinfo {pages} {053910}
  (\bibinfo {year} {2015})}\BibitemShut {NoStop}%
\bibitem [{\citenamefont {Bhoi}\ \emph {et~al.}(2021)\citenamefont {Bhoi},
  \citenamefont {Jang}, \citenamefont {Kim},\ and\ \citenamefont
  {Kim}}]{bhoi_2021_JAP}%
  \BibitemOpen
  \bibfield  {author} {\bibinfo {author} {\bibfnamefont {B.}~\bibnamefont
  {Bhoi}}, \bibinfo {author} {\bibfnamefont {S.-H.}\ \bibnamefont {Jang}},
  \bibinfo {author} {\bibfnamefont {B.}~\bibnamefont {Kim}},\ and\ \bibinfo
  {author} {\bibfnamefont {S.-K.}\ \bibnamefont {Kim}},\ }\href@noop {}
  {\bibfield  {journal} {\bibinfo  {journal} {Journal of Applied Physics}\
  }\textbf {\bibinfo {volume} {129}},\ \bibinfo {pages} {083904} (\bibinfo
  {year} {2021})}\BibitemShut {NoStop}%
\bibitem [{\citenamefont {Zhang}\ \emph {et~al.}(2019)\citenamefont {Zhang},
  \citenamefont {Ding}, \citenamefont {Zhou}, \citenamefont {Xu},\ and\
  \citenamefont {Jin}}]{XZhang_2019_PRL}%
  \BibitemOpen
  \bibfield  {author} {\bibinfo {author} {\bibfnamefont {X.}~\bibnamefont
  {Zhang}}, \bibinfo {author} {\bibfnamefont {K.}~\bibnamefont {Ding}},
  \bibinfo {author} {\bibfnamefont {X.}~\bibnamefont {Zhou}}, \bibinfo {author}
  {\bibfnamefont {J.}~\bibnamefont {Xu}},\ and\ \bibinfo {author}
  {\bibfnamefont {D.}~\bibnamefont {Jin}},\ }\href@noop {} {\bibfield
  {journal} {\bibinfo  {journal} {Physical Review Letters}\ }\textbf {\bibinfo
  {volume} {123}},\ \bibinfo {pages} {237202} (\bibinfo {year}
  {2019})}\BibitemShut {NoStop}%
\bibitem [{\citenamefont {Ihn}\ \emph {et~al.}(2020)\citenamefont {Ihn},
  \citenamefont {Lee}, \citenamefont {Kim}, \citenamefont {Yim},\ and\
  \citenamefont {Kim}}]{Ihn_2020_PRB}%
  \BibitemOpen
  \bibfield  {author} {\bibinfo {author} {\bibfnamefont {Y.~S.}\ \bibnamefont
  {Ihn}}, \bibinfo {author} {\bibfnamefont {S.-Y.}\ \bibnamefont {Lee}},
  \bibinfo {author} {\bibfnamefont {D.}~\bibnamefont {Kim}}, \bibinfo {author}
  {\bibfnamefont {S.~H.}\ \bibnamefont {Yim}},\ and\ \bibinfo {author}
  {\bibfnamefont {Z.}~\bibnamefont {Kim}},\ }\href@noop {} {\bibfield
  {journal} {\bibinfo  {journal} {Physical Review B}\ }\textbf {\bibinfo
  {volume} {102}},\ \bibinfo {pages} {064418} (\bibinfo {year}
  {2020})}\BibitemShut {NoStop}%
\bibitem [{\citenamefont {Boventer}\ \emph {et~al.}(2020)\citenamefont
  {Boventer}, \citenamefont {D{\"o}rflinger}, \citenamefont {Wolz},
  \citenamefont {Mac{\^e}do}, \citenamefont {Lebrun}, \citenamefont
  {Kl{\"a}ui},\ and\ \citenamefont {Weides}}]{Boventer_2020_PRR}%
  \BibitemOpen
  \bibfield  {author} {\bibinfo {author} {\bibfnamefont {I.}~\bibnamefont
  {Boventer}}, \bibinfo {author} {\bibfnamefont {C.}~\bibnamefont
  {D{\"o}rflinger}}, \bibinfo {author} {\bibfnamefont {T.}~\bibnamefont
  {Wolz}}, \bibinfo {author} {\bibfnamefont {R.}~\bibnamefont {Mac{\^e}do}},
  \bibinfo {author} {\bibfnamefont {R.}~\bibnamefont {Lebrun}}, \bibinfo
  {author} {\bibfnamefont {M.}~\bibnamefont {Kl{\"a}ui}},\ and\ \bibinfo
  {author} {\bibfnamefont {M.}~\bibnamefont {Weides}},\ }\href@noop {}
  {\bibfield  {journal} {\bibinfo  {journal} {Physical Review Research}\
  }\textbf {\bibinfo {volume} {2}},\ \bibinfo {pages} {013154} (\bibinfo {year}
  {2020})}\BibitemShut {NoStop}%
\bibitem [{\citenamefont {Kalinikos}\ and\ \citenamefont
  {Slavin}(1986)}]{kalinikos_1986_JPC}%
  \BibitemOpen
  \bibfield  {author} {\bibinfo {author} {\bibfnamefont {B.}~\bibnamefont
  {Kalinikos}}\ and\ \bibinfo {author} {\bibfnamefont {A.}~\bibnamefont
  {Slavin}},\ }\href@noop {} {\bibfield  {journal} {\bibinfo  {journal}
  {Journal of Physics C: Solid State Physics}\ }\textbf {\bibinfo {volume}
  {19}},\ \bibinfo {pages} {7013} (\bibinfo {year} {1986})}\BibitemShut
  {NoStop}%
\end{thebibliography}%

\end{document}